\documentclass[a4paper]{spie}  %>>> use this instead for A4 paper
\usepackage[]{graphicx}

\title{A practical system for X-ray Interferometry} 

\author{R. Willingale
\skiplinehalf
Department of Physics and Astronomy, University of Leicester,\\
University Road, Leicester, LE1 7RH, UK}

\authorinfo{E-mail: rw@star.le.ac.uk} 

\begin{document} 
  \maketitle 

%%%%%%%%%%%%%%%%%%%%%%%%%%%%%%%%%%%%%%%%%%%%%%%%%%%%%%%%%%%%% 
\begin{abstract}
X-ray interferometry has the potential to provide imaging at
ultra high angular resolutions of 100 micro arc seconds or better.
However, designing a practical interferometer which fits within a
reasonable envelope and that has sufficient collecting area to deliver such a
performance is a challenge. A simple system which can be built using
current X-ray optics capabilities and existing detector technology is described.
The complete instrument would be ~20 m long and ~2 m in diameter.
Simulations demonstrate that it has the sensitivity to provide high quality
X-ray interferometric imaging of a large number of available targets.
\end{abstract}

%>>>> Include a list of keywords after the abstract 


%%%%%%%%%%%%%%%%%%%%%%%%%%%%%%%%%%%%%%%%%%%%%%%%%%%%%%%%%%%%%
\section{INTRODUCTION}
\label{sect:intro}  % \label{} allows reference to this section
Interferometry is the primary method of imaging in radio astronomy
and is now providing high angular resolution in the optical band
\cite{baldwin}\cite{monnier}.
Because the wavelengths of X-rays are so small extremely high angular
resolutions should be possible using modest baselines providing
an X-ray interferometer can be built with sufficient precision.

In 2000 Cash et al. \cite{cash} reported the detection of X-ray fringes using
a simple grazing incidence interferometer utilising four flat mirrors.
Their prototype instrument had a baseline of just one millimetre and
gave fringes at 1.25 keV (wavelength 10 \AA)
equivalent to an angular resolution at source of $\sim0.1$ arc seconds.
More recently the spatial coherence of X-rays from a synchrotron source,
wavelength 1 \AA, has been measured by Suzuki \cite{suzuki}
using a two-beam interferometer with prism optics. The angular
resolution obtained was $~0.02$ arc seconds using an effective
baseline of $~0.3$ mm.  Conventional grazing incidence
X-ray telescopes like the Chandra Observatory \cite{weisskopf}
can achieve sub-arc second angular resolution but the performance
is a long way short of the diffraction limit despite the incredible
tolerances achieved in manufacturing the mirror surfaces.
It is tempting to assume that the precision required for
interferometry with very high angular resolution is beyond the reach of
modern technology but Cash et al. \cite{cash} have demonstrated this is not the case.
Generating a simple two-source fringe pattern is possible using
available flat mirrors and increasing the baseline does not require
a pro rata increase in precision.

The challenge is to build an X-ray interferometer with a collecting area
large enough to provide good statistics in the detected fringes
while at the same time making the instrument compact and reasonably 
straightforward to construct. Ideally
the dimensions of the instrument should be driven by the upper limit set
for the baseline separation
rather than being
dictated by the geometry required for the mirrors and detectors.

\section{The diffraction limit}

Using an unobstructed circular aperture of diameter $2R=D$ the diffraction
limited point spread
function is expected to be Airy's disk, and Rayleigh's criterion
gives a diffraction limited resolution of $\Delta\theta=1.22\lambda/D$
where $\lambda$ is the wavelength of the radiation. If $\Delta\theta$ is
100 $\mu$as at 2 keV, then $\lambda=6.2$ \AA\/ and $D\approx1.3$ m, a modest
aperture about twice the diameter of a single XMM-Newton module
($D=0.7$ m) or slightly larger than the largest shell in the Chandra
telescope ($D=1.2$ m).
Assuming a detector resolution of $\Delta y=10$ $\mu$m in the focal plane
the focal length must be $F\approx40$ km to give this angular resolution and
the cone angle of rays from the outer
edge of the aperture would be $2R/F=2\phi\approx7$ arc seconds corresponding
to an f-ratio of $\sim f/30000$.

In order to achieve diffraction limited imaging all the optical paths from the
aperture to the focus must be 
identical. For a thin lens operating in the visible this is accomplished
by retarding the wavefronts near the axis using a thickness of
dielectric. Similarly, a normal incidence mirror must be parabolic
to achieve the equivalent effect on reflection instead of refraction.
Neither of these is practical in the X-ray
regime but utilizing two grazing incidence reflections at
X-ray wavelengths the Wolter I (or II) configuration can provide
equal path lengths over a small annular aperture and can, in principle,
provide diffraction limited imaging over a small field of view.
However, a nest
of Wolter I surfaces, such as used in Chandra or XMM-Newton, cannot provide
diffraction limited imaging over the full aperture covered. If $R_{j}$ is
the radius at the join plane between the paraboloid and hyperboloid
surfaces the path difference introduced is $\Delta_{j}=R_{j}^{2}/2F$ which
increases as a function of $R_{j}$ and therefore
wavefront samples from different
Wolter surface pairs will not add in phase at the focus.

Even if the Wolter I surfaces could be manufactured with sufficient accuracy
the grazing angles required to meet the very large f-ratio would be just
a few
arc seconds and the resulting collecting area would be far too small to make it
practicable.

\section{Interferometric imaging}

\begin{figure}[!htb]
\begin{center}
\begin{tabular}{c}
\includegraphics[height=12cm,angle=-90]{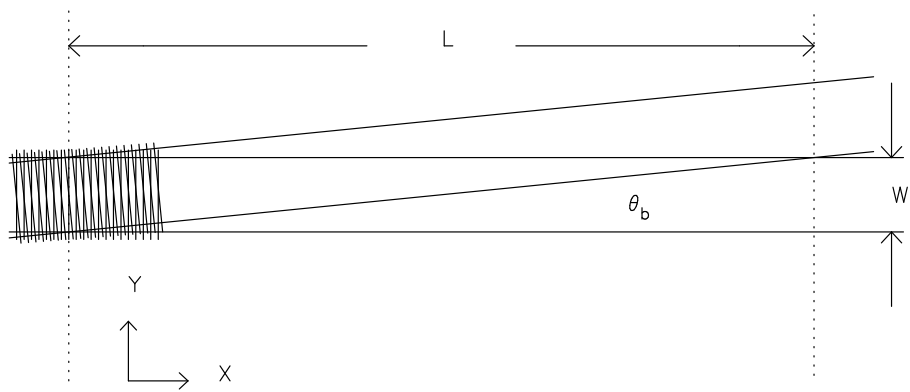}
\end{tabular}
\end{center}
\caption[fig1]{The geometry of two overlapping beams generating two-source
fringes.}
\label{fig1}
\end{figure}

Fig. \ref{fig1} shows the basic geometry needed to generate two-source
interference fringes in a wavefront splitting interferometer.
Parallel beams from two samples of the incident wave wavefronts
enter from the right and converge until they overlap to the left
creating a volume containing the interference fringes. Regardless of
how the beams are manipulated to the right, a length L, as shown,
is required to combine the beams and generate the fringes.

If the angle between the beams is $\theta_{b}$ and the wavelength is
$\lambda$ then the fringe spacing along the y-axis is

\begin{equation}
\label{eq1}
\Delta y=\frac{\lambda}{\theta_{b}}
\end{equation}

If the beam width is $W$ and the distance along the x-axis from
the position where the beams are separate to where the beams fully
overlap is $L$, then

\begin{equation}
\label{eq2}
\theta_{b}=\frac{W}{L}
\end{equation}

The number of fringes seen across the overlapping beams is given by

\begin{equation}
\label{eq3}
N_{f}=\frac{W}{\Delta y}
\end{equation}

Eliminating $\theta_{b}$ from equations \ref{eq1},\ref{eq2} and \ref{eq3}
the fringe separation and beam width are given by

\begin{equation}
\label{eq4}
\Delta y = \sqrt{\frac{\lambda L}{N_{f}}}
\end{equation}

\begin{equation}
\label{eq5}
W = \sqrt{\lambda LN_{f}}
\end{equation}

If we take $L=10$ m, $\lambda=10$ \AA\/ and $N_{f}=10$ then
$\Delta y = 30$ $\mu$m and $W=300$ $\mu$m.
The beam width $W$ is very small and in order to achieve a large collecting
area the depth of the beams along $Z$ (into the paper in Fig. \ref{fig1})
must be large and/or many identical systems must be operated in parallel.
The fringe spacing is small but can be resolved by current
X-ray imaging detectors. The situation can be improved by increasing
$L$ but we have already chosen a reasonably large distance
of $10$ m and because both $W$ and $\Delta y$ depend on $\sqrt{L}$ a
rather large increase is required to make a significant impact.
The angle between the beams is small, $\theta_{b}=6.2$ arc seconds.

\section{An X-ray interferometer}

\begin{figure}[!htb]
\begin{center}
\begin{tabular}{c}
\includegraphics[height=14cm,angle=-90]{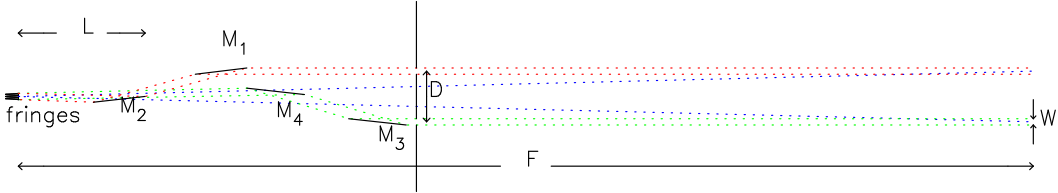}
\end{tabular}
\end{center}
\caption[fig2]{The four flat mirror configuration. The diagram is not
drawn to scale. The axial distance between $M_{2}$ and
$M_{3}$ is much less than the distance L and $F$ is much larger than
$L$. The vertical scale is
exaggerated so that the beam widths are visible.}
\label{fig2}
\end{figure}

Four flat mirrors can be used to take two samples of width $W$ and
separation $D=2R$ from the aperture and produce overlapping beams as shown
in Fig. \ref{fig1}.
Fig. \ref{fig2} shows the proposed four flat mirror configuration.
All four mirrors are set at
grazing incidence to provide high reflectivity. Operating in
the soft X-ray band 0.1-2.0 keV the grazing angles need to be
$\theta_{g}\approx 2^{\circ}$. If $M_{1}$ and $M_{3}$ are set at $\theta_{g}$
with respect to the x-axis then $M_{2}$ and $M_{4}$
must be set at a slightly smaller angle
$\theta_{g}-\theta_{b}/4$ so that the beams overlap to form fringes.
Since $\theta_{b}$ is very small compared to $\theta_{g}$, $M_{2}$ is
almost parallel to $M_{1}$ and the same is true for $M_{4}$ and $M_{3}$.
The effective focal length $F$ is much larger than $L$ and
is given by $\phi=\theta_{b}/2=\tan^{-1}(R/F)$.
The fringe spacing is then $\Delta y=F\delta \theta=F\lambda/D$
where $\Delta \theta$ is the diffraction limited angular resolution
for the baseline separation $D$ operating at wavelength $\lambda$.

The arrangement is similar to that used by Cash et al. \cite{cash} but
the front of $M_{2}$ is placed at distance
$L$ from the maximum overlap of the beams whereas in the Cash configuration this
mirror is at a distance $2L$ from the maximum overlap. The present
configuration reduces the overall length required by a useful factor of two.
The physical length of the system is much smaller than the focal length
as is the case for a Wolter II telescope.
If $W=300$ $\mu$m (see above) then the axial length of the mirrors
is only 8.6 mm if $\theta_{g}=2^{\circ}$.
The axial distance covered by the combination of $M_{1}$ and $M_{4}$
need only be $\sim25$ mm. The axial distance between $M_{1}$ and $M_{2}$
(or $M_{3}$ and $M_{4}$) is $D/(4\tan\theta_{g})\approx 7D$.

\subsection{A slatted mirror}

\begin{figure}[!htb]
\begin{center}
\begin{tabular}{c}
\includegraphics[height=12cm,angle=-90]{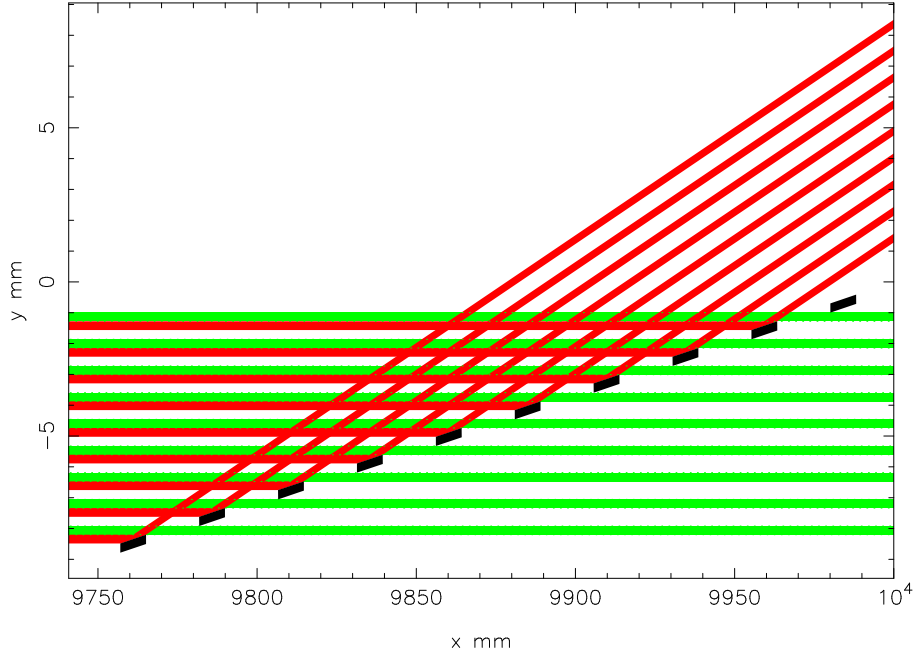}
\end{tabular}
\end{center}
\caption[fig3]{The beam paths using a slatted mirror}
\label{fig3}
\end{figure}

The collecting area afforded by a single four mirror arrangement
as illustrated in
Fig. \ref{fig2} is going to be small for any sensible depth of
mirror (along the z-axis into the plane of the paper). Furthermore
the mirrors required would be incredibly long and thin because
the axial length utilised is so small (8.6 mm, see above).
A major advantage of the four mirror configuration proposed here
and illustrated in Fig. \ref{fig2} is that
a series of parallel systems can be stacked together. In order to do this
the axial length of
mirrors $M_{1}$, $M_{3}$ and $M_{4}$ must be extended thereby
increasing the aperture widths and the mirror $M_{2}$ must be
split into a slatted mirror as shown in Fig. \ref{fig3}.

\begin{figure}[!htb]
\begin{center}
\begin{tabular}{c}
\includegraphics[height=15cm,angle=-90]{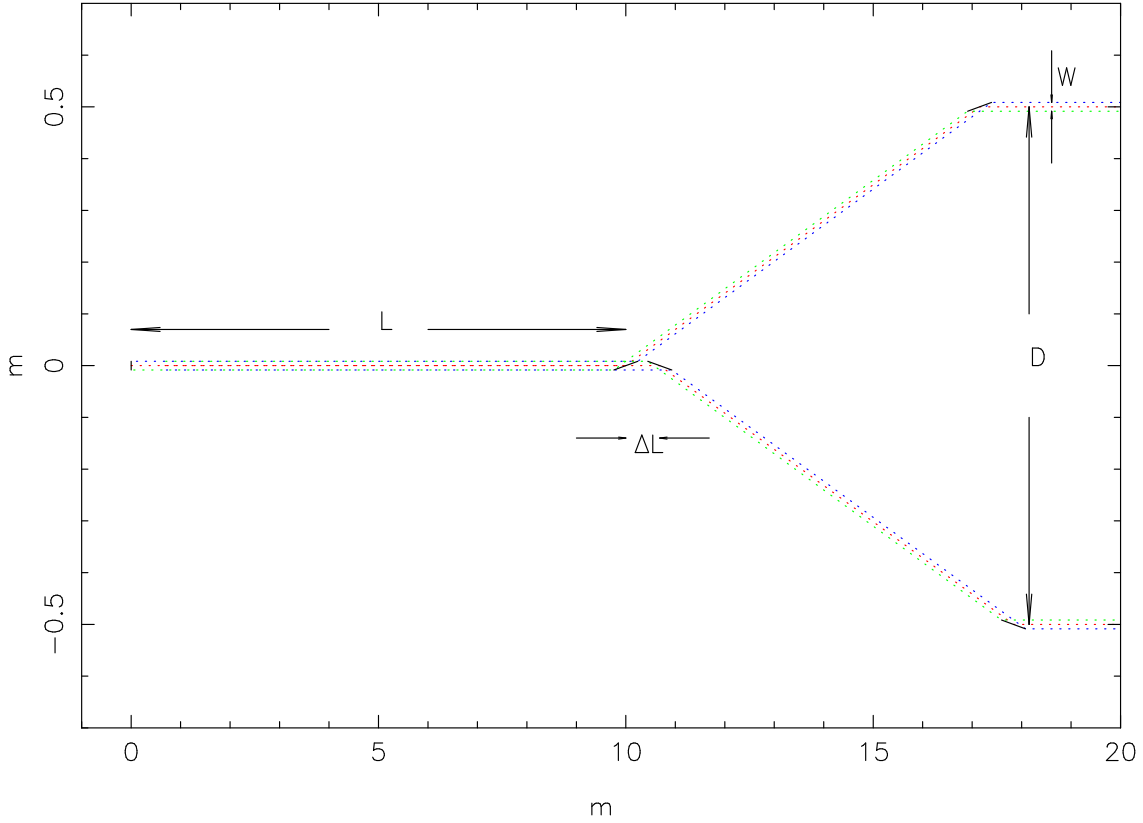}
\end{tabular}
\end{center}
\caption[fig4]{The layout of an X-ray interferometer using a slatted mirror
with 30 elements. Note that the vertical scale is expanded so that
the geometry of the beams is easier to discern.}
\label{fig4}
\end{figure}

Each slat is
a long thin mirror facet extending into the plane of the paper. The axial
width of a slat is the same as for the mirrors in Fig. \ref{fig2}.
The slats are spaced so that the beam from mirror $M_{4}$, to the right,
is broken into a series of beams of width $W$. Each slat mirror
reflects a fraction of the beam from mirror $M_{1}$ creating a second
set of beams.  Each pair of beams overlap to form
interference fringes. Providing there is not too much blocking from
support structure needed to hold the slatted mirror together
and provided the thickness of the mirror slats is the same order as
the beam width $W$ then about one third
of the flux collected by the apertures of $M_{1}$ and $M_{3}$
will form fringes. A slatted mirror with $\sim30$ slats will provide
an effective aperture width of $\sim1$ cm. The axial length for each
slat-gap pair will be $\sim26$ mm. Fig. \ref{fig4}
is a schematic diagram of the layout using a slatted mirror with 30 elements.
The length $\Delta L$ is the axial separation of $M_{1}$ and $M_{4}$ and
the baseline separation $D$ is the same for all the slat-gap pairs.

Such a slatted mirror is like a macroscopic transmission grating
with mirror facets on each line. It is likely that an optical
element of this form
could be manufactured using similar techniques to those
currently employed in the fabrication of
X-ray transmission or reflection gratings.
A slatted mirror with dimensions $500\times500$ mm combined
with 3 plane mirrors of the same size would provide
a total collecting area $\sim50$ cm$^{2}$.\footnote{The final effective
collecting area would depend on the X-ray reflectivity of the mirrors
and the efficiency of the detectors.}

\subsection{Path lengths}

If the mirrors in each arm
are set parallel the path length from the aperture at
$M_{3}$ to the detector plane is
\begin{equation}
\label{eq6}
P_{0}=(L+\Delta L+\frac{D}{2\cos 2\theta_{g}})(\frac{1}{\cos\theta}-1)
\approx (L+\Delta L+\frac{D}{2\cos 2\theta_{g}})\frac{\theta^{2}}{2}
\end{equation}
where $\theta$ is the off-axis angle of the source. This is the same
for both arms $M_{1}$-$M_{2}$ and $M_{3}$-$M_{4}$ and for all positions
across the detector.
When $M_{2}$ and $M_{4}$ are tilted by $\theta_{b}/4$ to produce
an overlap between the beams we get an extra
path contribution. The extra path lengths of the two arms are then:
\begin{equation}
\label{eq7}
P_{12}=L(\frac{1}{\cos(\theta_{b}/2)}-1)+y\sin(\theta_{b}/2)-
\frac{D}{2}\sin\theta
\end{equation}
\begin{equation}
\label{eq8}
P_{34}=(L+\Delta L)(\frac{1}{\cos(\theta_{b}/2)}-1)-y\sin(\theta_{b}/2)+
\frac{D}{2}\sin\theta
\end{equation}
where y is the position across the detector plane.
The path difference between the arms is:
\begin{equation}
\label{eq9}
\Delta=P_{12}-P_{34}=-\Delta L(\frac{1}{\cos(\theta_{b}/2)}-1)+
2y\sin(\theta_{b}/2)+D\sin\theta
\end{equation}

It is this path difference which give rise to the fringes. The first
term is fixed and of no consequence since it can be eliminated by
a small change in the position of $M_{1}$ or $M_{3}$.
Ignoring this small correction the coincidence point of the
interferemeter ($\Delta=0$) is given by
\begin{equation}
\label{eq10}
y=\frac{-D\sin\theta}{2\sin(\theta_{b}/2)} \approx-F\theta
\end{equation}
Here we have taken the small angle approximation and substituted for the focal
length $F=R/\tan(\theta_{b}/2)\approx D/\theta_{b}$.
The interferometer behaves like a telescope of focal length $F$
with the coincidence point (centre of the fringe pattern) at the
expected position of a point source with off-axis angle $\theta$. The
negative sign represents the expected lateral inversion in the focal plane.

\subsection{The fringe pattern}

If we move away from the coincidence point the path difference $\Delta$
increases linearly with $y'=y+F\theta$
and we expect to observe cosine fringes.
Because the wavefronts of the two beams are broken up by the slatted mirror
we must use Fresnel diffraction theory to calculate the exact form of the
fringe pattern. If plane waves of wavelength $\lambda$
are incident on a slit of width $W$
and we are looking at the fringes at a distance $L$ from the
slit the dimensionless variable used in the Fresnel integrals is given
by

\begin{equation}
\label{eq11}
u=y'\sqrt{\frac{2}{\lambda L}}
\end{equation}

Substituting for $y'=W$ from equation \ref{eq5} we have $u_{0}=\sqrt{2N_{f}}$.
Since $N_{f}>1$ the scaled width of the slit $u_{0}$
is also $>1$ and we
must use the near field approximation (Fresnel diffraction) rather than
the far field limit (Fraunhofer diffraction).

We define limits $u_{1}=u-u_{0}/2$ and $u_{2}=u+u_{0}/2$.
The complex amplitude at a scaled displacement $u$ from the centre of the
beam is given by

\begin{equation}
\label{eq12}
A=C(u_{2})-C(u_{1})+i(S(u_{2})-S(u_{1}))
\end{equation}

where $C(u)$ and $S(u)$ are the Fresnel integrals

\begin{equation}
\label{eq13}
C(u)=\int_{0}^{u}\cos(\pi w^{2}/2)dw
\end{equation}
\begin{equation}
\label{eq14}
S(u)=\int_{0}^{u}\sin(\pi w^{2}/2)dw
\end{equation}

\begin{figure}[!htb]
\begin{center}
\begin{tabular}{c}
\includegraphics[height=7cm,angle=-90]{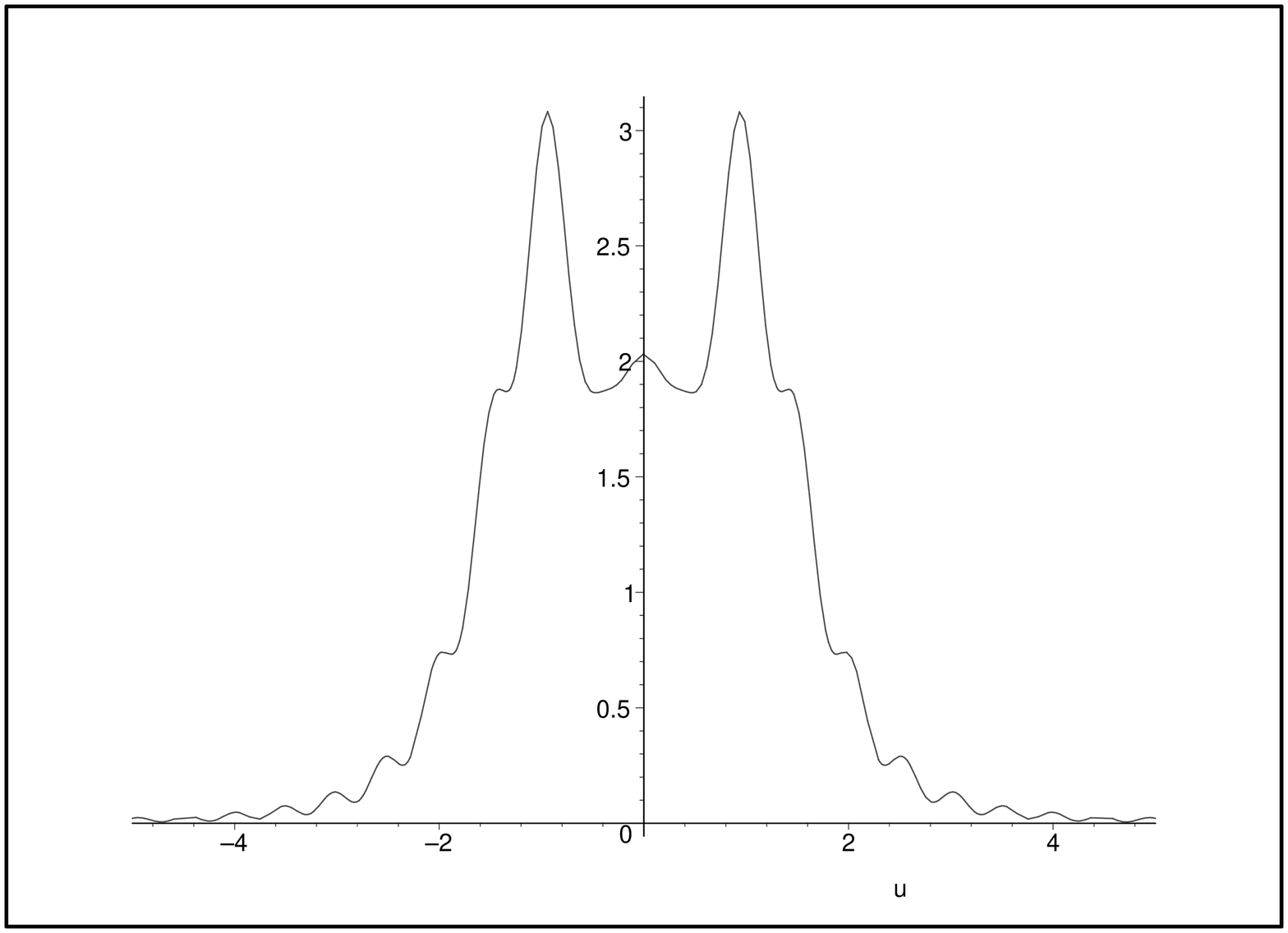}%
\hspace{0.5cm}
\includegraphics[height=7cm,angle=-90]{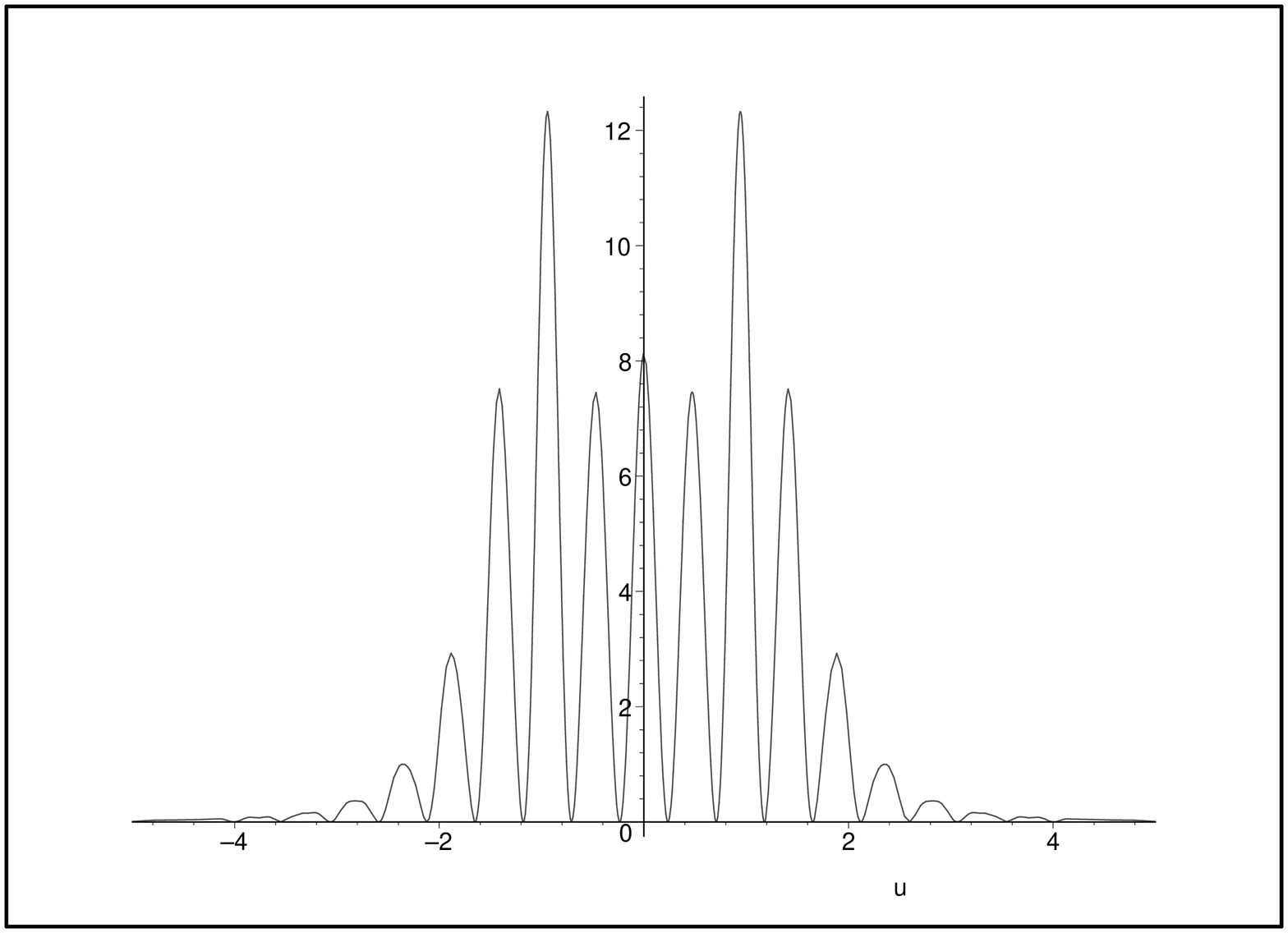}
\end{tabular}
\end{center}
\caption[fig5]{Left panel: The Fresnel diffraction profile of a
single beam. Right panel: The fringe pattern from one slat-gap pair.}
\label{fig5}
\end{figure}

The intensity expected is then given by $I=A A^{*}$.
Using the beam parameters from above $u_{0}=4.5$ and the intensity has the
profile shown in the left-hand panel of
Fig. \ref{fig5}. The geometric shadow of the edges
of the slit (a mirror slat or gap between slats) without diffraction are
expected at $u=\pm2.25$.

If the mirror slats and gaps are the same
size they will produce identical intensity profiles but because
they are tilted by $\theta_{b}$ with respect to each other
there is a phase difference between the beams which is a linear
function of $u$, $\delta=\pi u_{0}u$,
and the complex amplitude in the overlap region is then given by

\begin{equation}
\label{eq15}
A_{2}=A(1+\exp(i\pi u_{0} u))
\end{equation}

Again we can calculate the intensity profile in the same way
giving the fringe pattern plotted in right-hand panel of Fig. \ref{fig5}.
The intensity of the bright fringes
is modulated by the Fresnel diffraction profile shown in the left-hand
panel of Fig. \ref{fig5}.
The expected $N_{f}=10$ fringes are visible across the centre of the beam.
The edges of the beam spread into the geometric
shadow due to diffraction but there should be negligible interference
between adjacent slat-gap pairs.

\begin{figure}[!htb]
\begin{center}
\begin{tabular}{c}
\includegraphics[width=12cm,angle=-90]{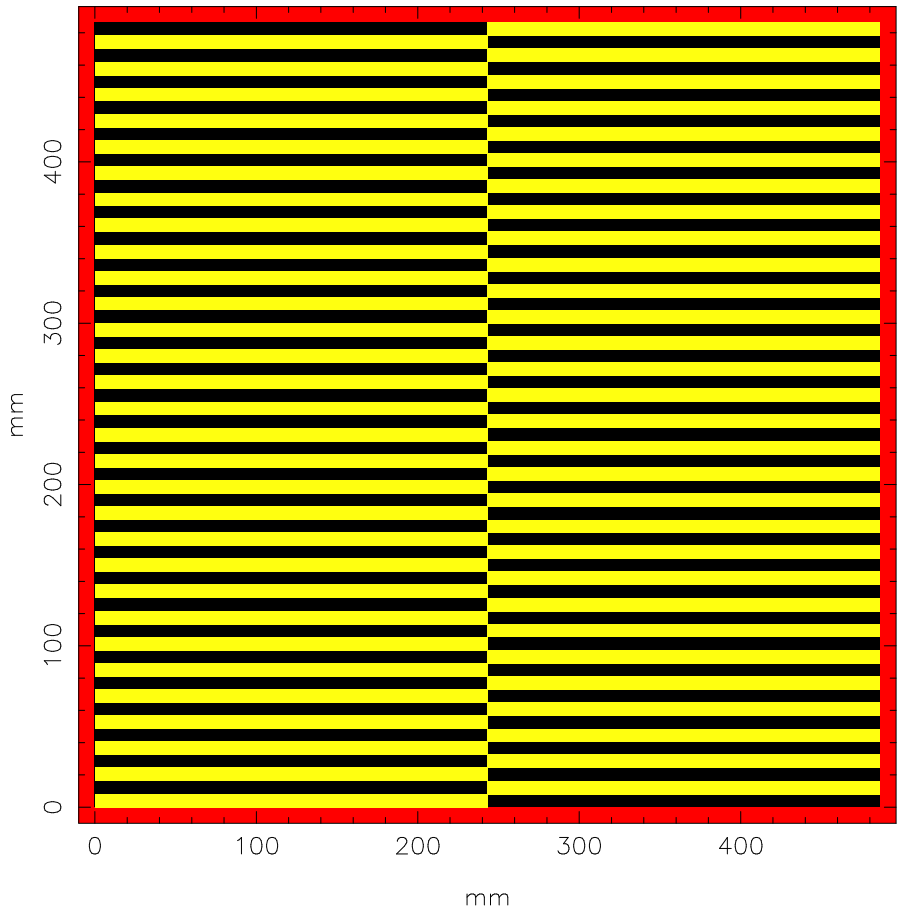}
\end{tabular}
\end{center}
\caption[fig6]{A slatted mirror with complementary halves.}
\label{fig6}
\end{figure}

\begin{figure}[!htb]
\begin{center}
\begin{tabular}{c}
\includegraphics[width=9cm,angle=-90]{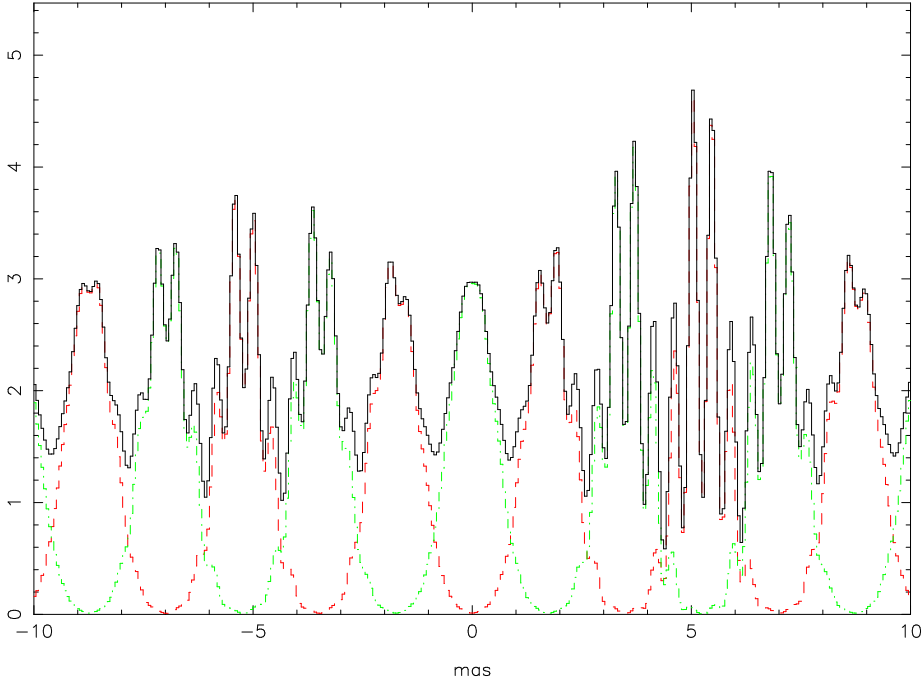}
\end{tabular}
\end{center}
\caption[fig7]{The fringe pattern expected from 2 sources
at $\pm5$ mas. The fringes from the two halves of the slatted mirror
are plotted as dash and dash-dot lines.}
\label{fig7}
\end{figure}

As the path difference $\Delta$ becomes comparable to the coherence length of
the X-rays the visibility of the fringes will decrease. If $E/\Delta E=N$ then
we expect to see $\sim N$ fringes across the entire pattern. If $N>>N_{f}$
then a continuation of
the fringes will be visible from slat-gap pairs adjacent to the 
coincidence position. However only $\sim$half the fringes will be detected
because of the gaps. These {\em missing} fringes can be recovered by
splitting the slatted mirror into two halves, reversing the
slat and gap positions in the second half. The pattern of slats
required is shown in Fig. \ref{fig6}. Combining the fringe patterns
from the two halves provides complete coverage of all $N$ fringes.
Fig. \ref{fig7} shows the fringe pattern expected from two sources
at $\pm5$ mas with $N_{f}\sim4$ and $N\sim10$. The slats introduce
a residual modulation but this can be completely removed during analysis of the
interferograms.

\subsection{Detecting the fringe pattern}

Working at
an X-ray wavelength of $10$ \AA\/ and using $L=10$ m the fringe spacing
for $N_{f}=10$ is $30$ $\mu$m. Such fringes could be resolved using
a CCD detector with the smaller pixel sizes currently available.
However the fringes exist through a very long volume in which the two
beams overlap. All 10 fringes
should be visible over an axial depth of $\sim L/10$.
If the detector is set at a grazing angle $\theta_{d}$ to the beam the
fringe spacing will be increased to

\begin{equation}
\label{eq16}
\Delta y'=\frac{\Delta y}{\sin\theta_{d}}
\end{equation}

If $\theta_{d}=5.7^{\circ}$ then the magnification factor will be 10
and the fringes will easily be resolved by current detector technology.
Unfortunately a detector operating at such a low grazing angle will have
a low efficiency. In order to take advantage of the magnification
a detector with high quantum efficiency operating at small grazing angles
would have to be developed.

When observing astronomical objects the X-ray flux will be broadband and
a detector with a moderate energy resolution will be required to detect
fringes. We require an energy resolution $E/\Delta E\geq N_{f}$ to
resolve the fringes at energy $E$ in a bandwidth $\Delta E$.
A CCD typically has $E/\Delta E\approx10$ at 0.6 keV increasing to
$\sim15$ at 1 keV and $\sim50$ at 7 keV. This is just adequate for our purpose.
However, the imaging and spectral response of a CCD is not well matched to the
requirement. High spatial resolution is only needed in 1-D (across the fringes)
and the sensitivity would be greatly improved if $E/\Delta E>100$.

\subsection{Simulation of one-dimensional imaging}

The interferometer illustrated in Fig. \ref{fig4} has been simulated
using the slatted mirror layout shown in Fig. \ref{fig6} and
the energy response of the XMM-Newton EPIC-MOS CCDs \cite{turner}.
In order to get good coverage on the u-axis (spatial frequency) four
parallel systems were used with D-spacings of 35, 105, 315 and 945 mm.
The corresponding effective focal lengths are 1.2, 3.7, 11.1 and 33.4 km.
Each system has a collecting area of $\sim20$ cm$^{2}$ in the
energy band 0.58-2.1 keV (using the reflectivity of gold and the
MOS detector efficiency). The $E/\Delta E$ of the detector provides
21 energy channels across this band. A total source flux equivalent
to 1 Crab gives 460000 counts in a 1000 second exposure. With
4 D-spacings and 21 energy channels a total of 84 interferograms
were recorded in a single exposure. The source distribution assumed
was a binary system consisting of an extended source and a point-like
companion.

\begin{figure}[!htb]
\begin{center}
\begin{tabular}{c}
\includegraphics[width=10cm,angle=-90]{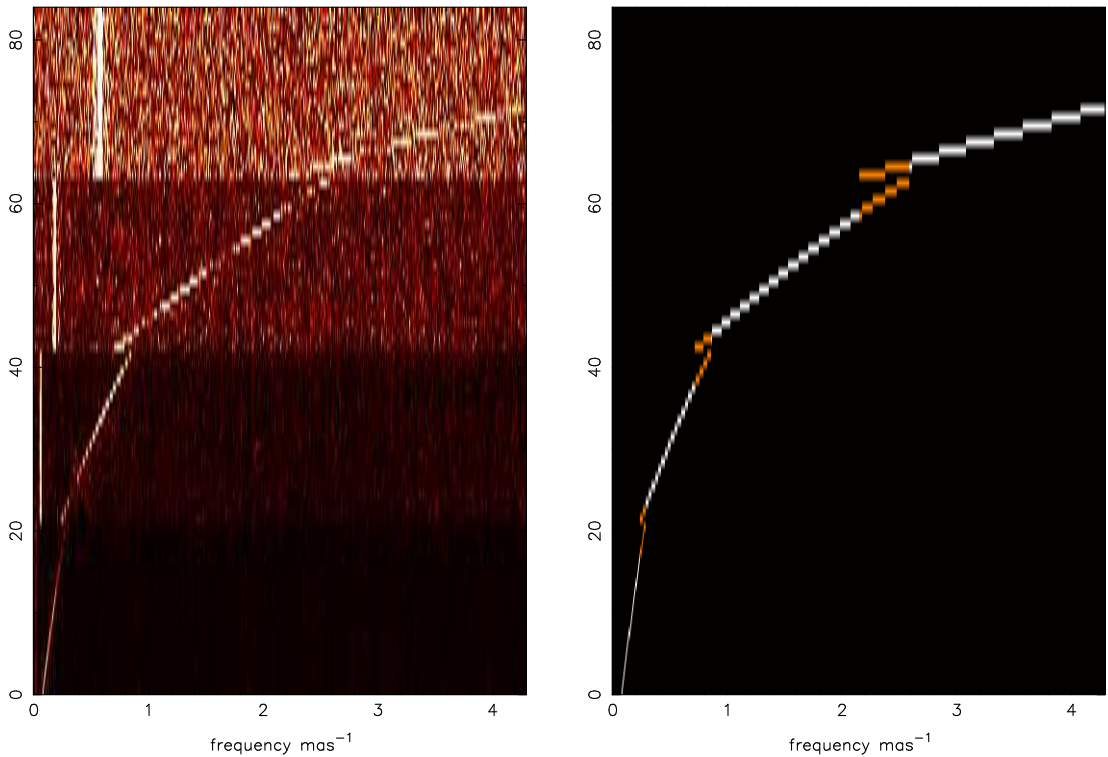}
\end{tabular}
\end{center}
\caption[fig8]{The left-hand panel shows the Fourier power spectra
of 84 interferograms. Each block of 21 corresponds to a given D-spacing.
The right-hand panel is the Fourier filter used to pick out the fringes.}
\label{fig8}
\end{figure}

\begin{figure}[!htb]
\begin{center}
\begin{tabular}{c}
\includegraphics[width=9cm,angle=-90]{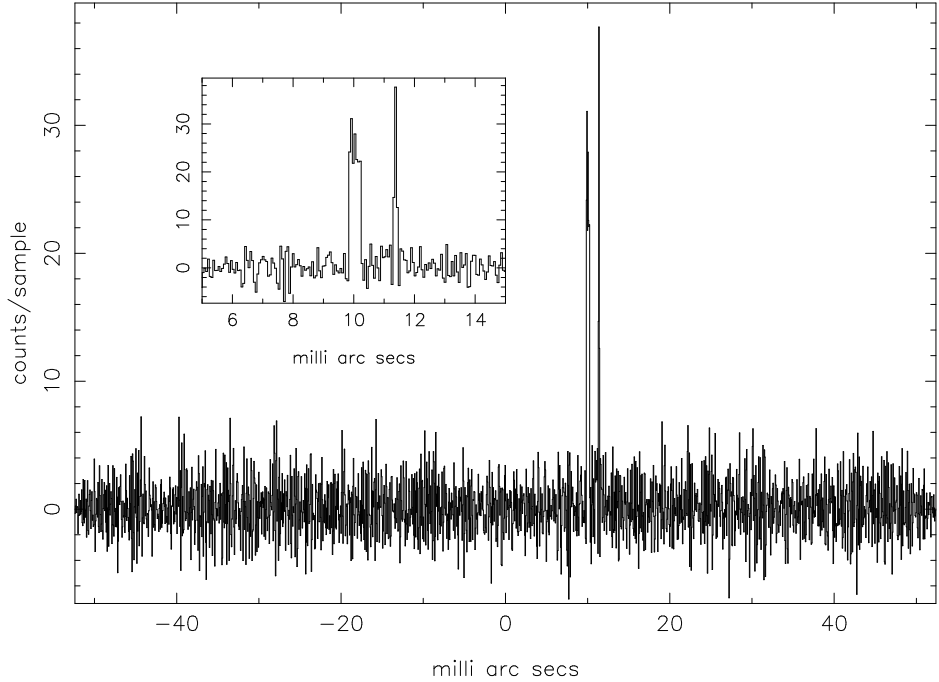}
\end{tabular}
\end{center}
\caption[fig9]{The reconstructed source distribution. The inset
shows an expanded detail of the significant peaks. The binary consists
of a resolved component with a point source companion.}
\label{fig9}
\end{figure}

Even with $\sim460000$ counts the count per fringe is very small and
it is impossible to see the fringes in the raw simulated data. However,
for each interferogram (1 energy channel and 1 D-spacing) the 
fringe spacing and expected number of fringes is known. It is therefore
possible to set up a Fourier filter that picks out the fringe
pattern from each interferogram. Fig. \ref{fig8} shows the
Fourier power spectra of the 84 interferograms and the Fourier filter
constructed to pick out the fringes.
The 4 blocks of 21 interferograms arise from the 4 D-spacings used.
Two peaks are visible in each power spectrum. The vertical white
lines to the left at
low frequency are the peaks from the modulation caused by the slats. These are
completely removed by the filtering. The white patches
to the right are the fringes.
The visibility of the fringes varies as the frequency increases because
of the structure of the binary source under observation. A top-hat
profile matched to the $E/\Delta E$ for each interferogram was used
to construct the filter.
There is some
overlap in the frequency coverage between the 4 D-spacings.
In a practical set up the overlap regions could provide a means of
eliminating phase errors between the 4 parallel optical systems.
Fig. \ref{fig9} shows the reconstruction of the source distribution.
The intensity is plotted as counts per 0.058 mas sample and there are
1800 samples across the field of view.
The rms noise level is 2.16 counts per sample and all the significant samples
are detected at $>12\sigma$. The total estimated count from the significant
samples is 430000 compared with the actual detected count of 462000 so
93\% of the original count has been successfully imaged.

In this simulation and reconstruction the source distribution was assumed
to be independent of energy and therefore all the detector energy
channels could be summed to produce the final image.
In reality this would not always be the case
and more D-spacings would be required to reconstruct a source with a
complex spatial-energy structure. To extend the imaging to 2-D more
exposures would be required at different roll angles about the pointing
axis to give
a reasonable coverage in the (u,v) plane. If this were achieved by running
several (5-10) identical systems simultaneously at different roll angles 
not only would 2-D imaging be provided but there would also
be a pro rata increase in the total collecting area.

The system performs in a similar way to a shadow mask camera \cite{sims}
but a fringe pattern is detected instead of a shadow mask pattern.
Each pixel at the detector is
multiplexed to many sky elements within the field of view and therefore the
sensitivity of the interferometer is dependant on the source distribution
or the number of significantly bright point sources in the field of view.
If the energy resolution of the detector were improved
the number of fringes $N\approx E/\Delta E$ across the pattern would
increase and the number of pixels multiplexed to a given sky position
would be larger. The total area of the Fourier plane covered by the
filter (Fig. \ref{fig8}) is  $\propto 1/N$ and if there are $N_{x}$ significant
unresolved sources in the field-of-view the signal-to-noise is
$\propto \sqrt(N)/N_{x}$.

\section{Tolerances, alignment and adjustment}

In the practical implementation of the interferometer we must consider:

\begin{itemize}
\item figure errors in the flat mirror surfaces
\item roughness of the mirror surfaces
\item angular alignment of the mirrors
\item positional placement of the mirrors
\item control of the difference between the path lengths in the two beams
\item pointing accuracy and stability
\end{itemize}

\subsection{Figure errors and surface roughness}

The mirrors must be flat enough so that incident plane wavefronts are
reflected with the minimum perturbation and remain plane. Figure errors
will introduce distortions in the wavefronts and shorter scale
surface roughness features will scatter some of the incident light into
scattering wings which will reduce the contrast of the fringes.
Fortunately, because the mirrors are operating at
grazing incidence, the effect of figure errors and surface roughness
in the mirror surfaces is reduced by a factor $\sin \theta_{g}$.
A surface
height error $h$ introduces a wavefront shift of $2h\sin\theta_{g}$. To
produce clean fringes the wavefronts must not be perturbed by greater than
$\sim\lambda/10$. So we have

\begin{equation}
\label{eq17}
h\le\frac{\lambda}{20 \sin \theta_{g}}
\end{equation}

If $\lambda=10$ \AA\/ and $\theta_{g}=3^{\circ}$ then we require $h<1$ nm.
A high quality optical flat has a specification of $\lambda/20$ (where
$\lambda=633$ nm) so even with the advantage of grazing incidence we need
very high quality mirrors to obtain clean fringes. The surface height error
of 1 nm
is equivalent to an axial gradient error over the width of 1 slat
of $\sim0.03$ arc seconds. Gradient errors over distances larger than the slat
width will destroy the register between the overlapping beams
produced by a single slat-gap pair and may introduce confusion between
adjacent slat-gap pairs. The angular width of the fringe separation is
$\Delta y/L\approx 3\times10^{-6}$ equivalent to 0.6 arc seconds as seen
from the mirrors. Therefore
gradient errors between the slats and over the full faces of the other mirrors
must be kept to this level.

First order perturbation theory gives the Total Integrated Scatter (TIS)
after two reflections as

\begin{equation}
\label{eq18}
TIS=2(\frac{4\pi \sigma \sin \theta_{g}}{\lambda})^2
\end{equation}

where $\sigma$ is the rms surface roughness. To get $TIS<0.1$ we require
$\sigma<3.5$ \AA\/ integrated over correlation lengths $<1$ mm on the
surface.

The quality of mirror surfaces required is high, similar to the best
X-ray mirrors used in instruments such as Chandra. However the surfaces
are flat rather than aspherics and therefore they should be significantly
easier to manufacture than Wolter type I surfaces. Manufacture of the
slatted mirror imposes a much higher level of difficulty since the mirror
surfaces must be flat but the substrate must be very thin and extra support
will be required to hold the slats together without introducing too
much blocking of the aperture.

\subsection{Adjustment and stability of the D-spacing}

The baseline separation $D$ is set by the distances $d_{12}$ and $d_{34}$
measured between the mirror centres,
$D=(d_{12}+d_{34})/\sin 2\theta_{g}$.
These distances can be reduced until
a minimum $D_{min}=4WN_{s}$ where $N_{s}$ is the number of slats in
mirror 2. If the mirrors are brought any closer
than this the outer edges of the mirrors 2 and 4 will start to be blocked and
fringes will only be seen for the central slat-gap pairs in mirror 2.
$D_{min}$ is the minimum baseline for which the full collecting area is
available.

The distances $d_{12}$ and $d_{34}$ must be
the same so that the path lengths for the two beams are identical.
To change the baseline both mirrors 1 and 3 must be moved.
An off-axis angle of $\theta$ introduces a path difference
$\Delta=D\sin\theta$ (see equation \ref{eq9}).
Thus adjustments of $d_{12}$ and $d_{34}$ are coupled to the
pointing direction $\theta$. If you imagine that mirror 3 is fixed
at position $d_{34}$ and rotation $\theta_{3}$ then
the path lengths can then be equalized either by adjusting
$d_{12}$ or by changing the pointing $\theta$. Of course this
coupling is why the phase and visibility of the fringes observed give us
information on the angular distribution of the source.

The total path difference between the two beams
must be less than the coherence length of the X-rays or the fringes
will disappear. A coherence length of 10 $\mu$m corresponds to a wave
train of 10000 cycles but this order of coherence will only be
seen in line radiation for which the radiative lifetime is relatively
large. In astronomical observations the source spectrum
is dominated by continuum radiation and the effective coherence length
will be determined by the energy resolution of the detector system.
Using CCD technology at $\sim1$ keV the wave trains will be just 10
cycles long. In this case we must control $d_{12}-d_{34}$ to order
10 nm or set $\theta<10\lambda/D$. If $D=10$ mm then $\theta<0.2$ arc
seconds.

\section{Conclusion}

We have described the design of an X-ray interferometer
that can fit into a tube $\sim20$ m long and $\sim2$ m diameter. The simplest
configuration which can provide full 1-D imaging consists of 4 optical units.
Each unit comprises 3 flat mirrors, a flat slatted mirror and an array of
CCD detectors. The combination gives a collecting area of $\sim 80$ cm$^{2}$
in the energy band 0.58-2.1 keV and an angular resolution of $\sim0.1$ mas
over a field of view 100 mas across.
A simple field of view containing $N_{x}\approx 10$ significant 
unresolved sources with a total flux equivalent to 1 Crab can be imaged
to a sensitivity $>5\sigma$ in 500 seconds.
The system can provide imaging in 2-D by making exposures
at different roll angles. About 40 units (possibly packed into the same tube)
running simultaneously with different D-spacings and
roll angles could provide good coverage of the (u,v) plane and because of
the 10-fold increase in collecting area the same sensitivity would be
achieved in $\sim150$ seconds using CCDs or detectors with a similar
performance.
If the detector energy resolution could be improved by a factor of
10 while retaining a spatial resolution of $\sim10\mu$ m in 1-D then the
same 2-D imaging sensitivity could be achieved in $\sim50$ seconds.

The interferometer requires a high quality slatted mirror the specifications
of which are challenging but probably well within current manufacturing
capability.  However, such a mirror has not been produced and needs to be
developed.

The system is similar to the MAXIM periscope configuration,
the tolerances of which are described by Shipley et al. \cite{shipley}.
An equivalent study of the tolerances and tradeoffs is
required for the present concept to further the design of the optical bench
needed.

The introduction of a slatted mirror dramatically
reduces the total distance required between the primary mirrors
that define the baseline separation and the detector system
and 0.1 mas imaging can be achieved without the requirement for two
free-flying spacecraft as considered for the MAXIM Pathfinder
\cite{gendreau} or the many free flying spacecraft required
for the all-up MAXIM configuration \cite{lieber}.
Each unit of 4 mirrors detects two-source fringes and the combination of
several such units provides imaging by aperture synthesis in the same way as
conventional interferometers used in the radio and optical bands.
This is rather different from the all-up MAXIM approach
in which many mirror segments are used to produce a complex interferogram
which is much closer to a conventional image. As a first step
towards ultra high angular resolution X-ray astronomy the present
scheme is more compact and easier to implement than the MAXIM
configurations considered thus far.

%-------------
%%%%%%%%%%%%%%%%%%%%%%%%%%%%%%%%%%%%%%%%%%%%%%%%%%%%%%%%%%%%%
%\acknowledgments     %>>>> equivalent to \section*{ACKNOWLEDGMENTS}       
 
%%%%%%%%%%%%%%%%%%%%%%%%%%%%%%%%%%%%%%%%%%%%%%%%%%%%%%%%%%%%%
%%%%% References %%%%%

\end{document}